\documentclass{nature}
\usepackage{amsmath}
\usepackage{graphicx}
\usepackage{float}
\usepackage{color}

\makeatletter
\let\saved@includegraphics\includegraphics
\AtBeginDocument{\let\includegraphics\saved@includegraphics}
\renewenvironment*{figure}{\@float{figure}}{\end@float}
\makeatother

\title{Electric-Field Control of the Interlayer Exchange Coupling for Magnetization Switching}

\author{Shehrin Sayed$^{1,a}$, Cheng-Hsiang Hsu$^1$, Niklas Roschewsky$^2$, See-Hun Yang$^3$, \& Sayeef Salahuddin$^{1,b}$}

\begin{document}

\maketitle

\begin{affiliations}
 \item Department of Electrical Engineering and Computer Science, University of California, Berkeley, CA 94720, USA,
 \item Department of Physics, University of California, Berkeley, CA 94720, USA, and
 \item IBM Research - Almaden, San Jose, California 95120, USA.\\
 {\footnotesize $^a$ ssayed@berkeley.edu}\\
 {\footnotesize $^b$ sayeef@berkeley.edu}
\end{affiliations}

\begin{abstract}
We propose an electric-field-controlled mechanism for magnetization switching assisted solely by the interlayer-exchange coupling (IEC) between the fixed and the free magnets, which are separated by two oxide barriers sandwiching a spacer material known for exhibiting large IEC. The basic idea relies on the formation of a quantum-well (QW) within the spacer material and controlling the transmission coefficient across the structure with an electric-field via the resonant tunneling phenomena. Using non-equilibrium Green's function (NEGF) method, we show that the structure can exhibit a bias-dependent oscillatory IEC that can switch the free magnet to have either a parallel or an antiparallel configuration with respect to the fixed magnet, depending on the sign of the IEC. Such bi-directional switching can be achieved with the same voltage polarity but different magnitudes. With proper choice of the spacer material, the current in the structure can be significantly reduced. Due to the conservative nature of the exerted torque by the IEC, the switching threshold of the proposed mechanism is decoupled from the switching speed, while the conventional spin-torque devices exhibit a trade-off due to the non-conservative nature of the exerted torque.
\end{abstract}

\section{Introduction}


Magnetization switching using current-induced spin-transfer torque (STT) \cite{Berger1996,Slonczewski1996} has attracted increasing interest for non-volatile memory technologies like magnetoresistive random access memory \cite{Chappert2007, Apalkov2016, Kawahara2012}. However, the large current density required for STT switching is limiting the technological advancement in terms of the energy efficiency and bit density. Recently, there is a growing interest in the voltage or electric-field controlled switching mechanisms \cite{Matsukura2015, Newhouse-Illige2017, Wang2018, Cherifi2014, Heron2014} as a possible solution to address the issues involving memory bandwidth and high-power consumption \cite{Wang2018, Salahuddin2018}.

\textit{Summary of key contributions.} In this paper, we propose a composite structure that can enable electric-field controlled magnetization switching assisted solely by the interlayer exchange coupling (IEC) between the fixed and the free magnets. The magnets are separated by two tunnel barriers sandwiching a thin layer of a spacer material that exhibits large IEC. The electric-field control of IEC relies on the formation of a quantum-well (QW) within that spacer layer with discrete energy states positioned above the equilibrium Fermi level, which enables electric-field induced modulation of the transmission coefficient between the two magnets via the resonant tunneling phenomena. We use Non-Equilibrium Green's Function (NEGF) method \cite{Datta1995} to show that a sizable bias-dependent oscillatory IEC could be induced which, in turn, could switch the free magnet to have either a ferromagnetic (F) or an antiferromagnetic (AF) configuration with respect to the fixed magnet, depending on the sign of the IEC. The configuration will be retained once the electric-field is removed because the barriers suppress the equilibrium IEC. 

Switching in both directions could be possible for the same bias-polarity, but different magnitudes above the switching threshold, which is different from the existing electrical switching mechanisms. We use a coupled Landau-Lifshitz-Gilbert (LLG) equation \cite{Victora2005, Camsari2016} to show that the switching threshold of the electric-field controlled IEC energy is independent of the Gilbert damping and same for magnets with in-plane and perpendicular anisotropies. We argue that the proposed structure can be optimized to lower both the current in the structure and the operating voltage, while getting a sizable bias-dependent IEC for magnetization switching. On the other hand, the switching speed is inversely proportional to the Gilbert damping and slower for a perpendicular magnet as compared to an in-plane magnet. These unique features of the proposed structure suggest that devices with significantly improved energy-delay product can be designed.

\begin{figure}
	\centering
	\includegraphics[width=0.95\textwidth]{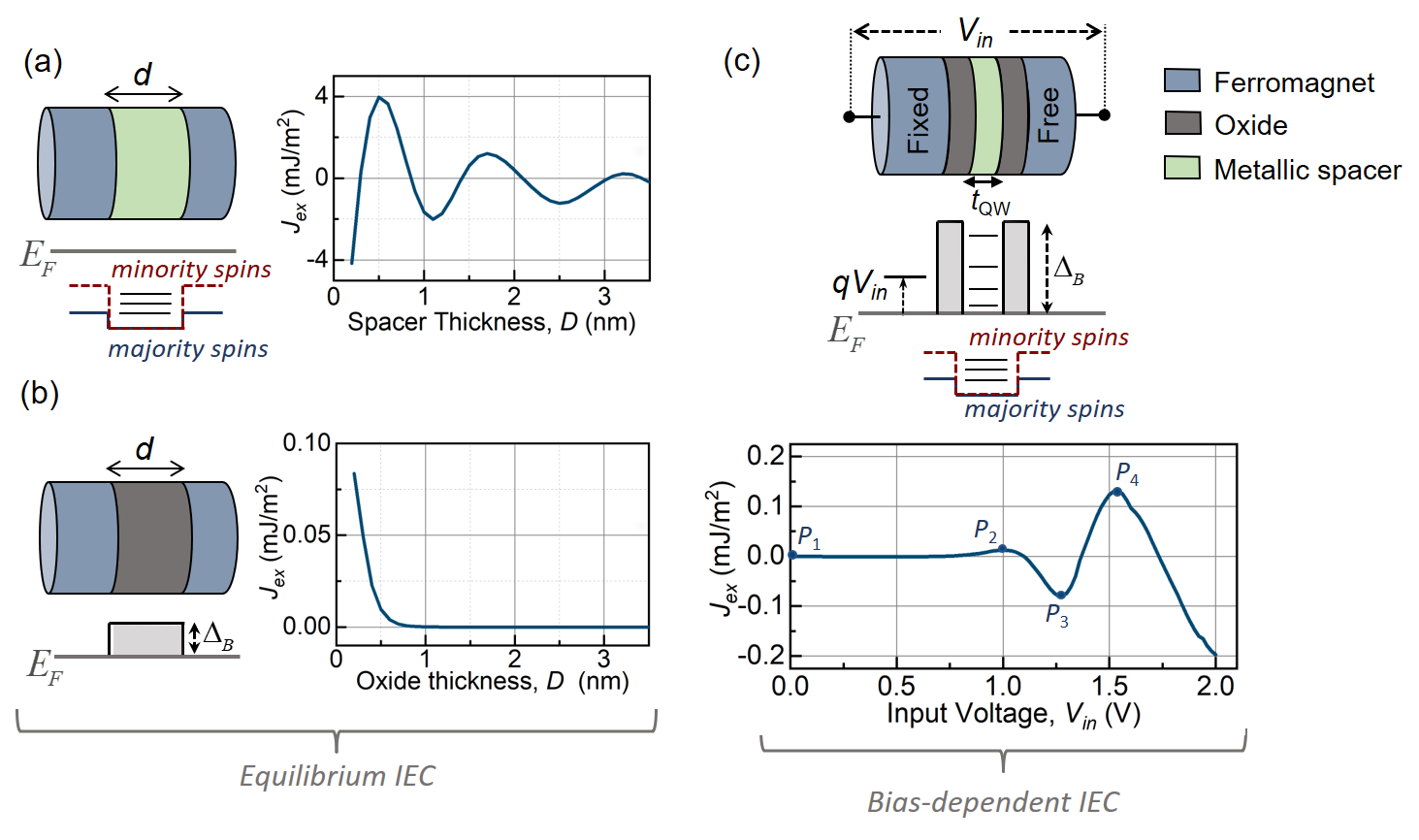}
	\caption{(a) Oscillatory interlayer exchange coupling (IEC) through a metallic spacer as a function of the spacer thickness $D$, which is attributed to the spin-dependent quantum-well seen by the majority and minority spins below the equilibrium Fermi level $E_F$. (b) Thin oxide exhibits non-negligible IEC while a thicker oxide diminish it. (c) A structure with double barrier sandwiching a metallic spacer which forms additional QW states above $E_F$ and the transparency between the magnets can be controlled using an electric-field via resonant tunneling phenomena. This structure will exhibit a bias-dependent oscillation in IEC that grows in magnitude. Barrier heights are 0.7 eV and widths are 1 nm each. Spacer thickness is 0.8 nm.}\label{1}
\end{figure}

\section{Electric-field control of the interlayer exchange coupling (IEC)}
\textit{Basic mechanism.} Our prediction is based on the well-established phenomenon that two ferromagnets (FM) separated by a metallic \cite{Parkin1991,Petroff1991,Fullerton1993,Bennett1990} or a non-metallic \cite{Toscano1992,PhysRevLett.91.037207} spacer layer prefers either a F or an AF configuration at equilibrium, depending on the sign of the IEC. The sign of the IEC oscillates periodically as a function of the spacer thickness (see Fig. \ref{1}(a)), due to the quantum-interference by the majority and the minority spins in the spacer as they see different QW like potential profiles below the equilibrium Fermi level $E_F$ formed at the magnetic interfaces \cite{Slonczewski1989, Bruno1995, Haney2009, Mathon1993}. On the other hand, a thick oxide barrier significantly reduces such IEC (see Fig. \ref{1}(b)) and the two FMs do not have a preferential configuration at equilibrium.

We introduce two oxide barriers at the two magnetic interfaces of the structure in Fig. \ref{1}(a) (see Fig. \ref{1}(c)), which form a QW with discrete energy states above $E_{F}$. At these discrete states, transmission coefficients are high as seen for the resonant tunneling diodes \cite{Chang1974}. These states can be probed by tuning the contact electrochemical potentials ($\mu_{1,2}$) with an electric-field across the structure. Subsequently, the two FMs feel a sizable IEC due to a spin-dependent interference by the filled states within the spacer, similar to the discussion in Fig. \ref{1}(a). We argue that the spin-dependent interference could be constructive or destructive depending on the electric-field controlled transmission coefficient, giving rise to a bias-dependent oscillation in IEC (see Fig. \ref{1}(c)). The IEC strength is also expected to grow with increasing electric-field as the number of occupied QW states increases. 

\textit{Model.} We have analyzed the IEC between two magnets based on parameters calculated using the Non-Equilibrium Green's Function (NEGF) method. The IEC energy per unit area is usually taken as the difference of energy density change across the QW ($\Delta E$) between the ferromagnetic (F) and the antiferromagnetic (AF) configurations \cite{Bruno1995, Haney2009, Slonczewski1989}, as given by
\begin{equation}
\label{IEC}
{J_{ex}} = \Delta E_F - \Delta E_{AF},
\end{equation}
where the change in the energy density across the QW is calculated as \cite{Slonczewski1989} $\Delta E = \int_{-\infty}^{+\infty} E \,\Delta n_s\, dE$, with $\Delta n_s=n_s(d)-n_s(0)$ being the change in the spin density across the QW. Note that the magnetization easy-axis is along $z$-direction in this discussion.

We calculate $n_s$ using the NEGF method using a single band effective mass 1D tight-binding Hamiltonian is used here for calculations. We assume that the structure under consideration is spatially uniform along the transverse directions and transverse modes are nearly decoupled so that transport can be analyzed with 1D Hamiltonian for every mode. The spin density is given by
\begin{equation}
\label{IEC}
n_s = \frac{1}{2\pi}\int D_0\;{\mathop{\rm Re}\nolimits} \left[ {{\rm{Tr}}\left[ { \sigma_z G^n} \right]} \right]\;d\epsilon_T,
\end{equation}
where $\epsilon_T=\dfrac{\hbar^2 \left(k_y^2+k_z^2\right)}{2m^*}$ is the energy along the transverse plane $y$-$z$, $D_0$ is the 2D density of states on the transverse plane, $\sigma_z$ is the $z$-Pauli matrix, and $G^n$ is the correlation function. We approximate the $D_0$ with the 2D density of states of the bulk given by ${m^*}/{\pi\hbar^2}$ where $m^*$ is the effective mass and $\hbar=h/2\pi$.

The correlation function is obtained as
\begin{equation}
\label{IEC}
G^n=\sum_j\;f_j\,A_j,
\end{equation}
where $f_j=1/\left(1+\exp{\left(\left(E-\mu_j\right)/k_BT\right)}\right)$ is the occupation factor of $j^\text{th}$ contact with contact electrochemical potential $\mu_j$, $A_j=G^R \, \Gamma_j \left(G^R\right)^\dagger$ is the spectral function with $\Gamma_j$ being the broadening function of the $j^\text{th}$ contact, $G^R = \left[EI-H-\Sigma\right]^{-1}$ is the Green's function, $\Sigma$ is the total self-energy of the contacts, and $H$ is the Hamiltonian of the structure. We assume that the voltage applied across the two-terminal structure mostly drops across the two oxide barriers. The details of the NEGF based calculations can be found in the supplementary information. In this paper, positive and negative IEC indicates AF and F configurations respectively.

\textit{Materials and Structure}. An ideal material for the spacer of the proposed structure in Fig. \ref{1}(c) could be the transition metals e.g. Rh, Ru, Ir, Re, and Cu \cite{Parkin1991, Yakushiji2017, Dinia1998}, which exhibit large IEC strengths at equilibrium in a geometry shown in Fig. \ref{1}(a). The strength of the equilibrium IEC across a metallic spacer in Fig. \ref{1}(a) depends on the shape of the spin-dependent QW, which is determined by the mismatch of the $d$-electron bands between the FM and the metallic spacer material \cite{Mathon1993, Parkin1991}, growth condition, and hybridization at the interfaces \cite{Dinia1998}. In this paper, we have calibrated the spacer parameters for NEGF calculations such that the first AF IEC strength in the structure in Fig. \ref{1}(a) is around $\sim4$ mJ/m$^2$, comparable to a Co$|$Ru system \cite{Parkin1991}. We have set the spacer thickness to 0.8 nm. 

The barrier heights in Fig. \ref{1}(c) were set around 0.7 eV, as typically observed for MgO \cite{Schleicher2014, Yuasa2004}. The barrier thicknesses were set to 1 nm each. It has been discussed both theoretically \cite{Yang2010} and experimentally \cite{Faure-Vincent2002, Chiang2009} that equilibirum IEC across such a thick oxide is negligible. A non-negligible equilibrium value could be seen for very thin oxide since the transmission coefficient could be due to the tunneling effect \cite{Faure-Vincent2002, Chiang2009}. Bias-dependent IEC through a reasonably thick single oxide barrier has been discussed theoretically via high voltage tunneling \cite{You1999, Tang2009} and experimentally via mobile oxygen vacancies \cite{Newhouse-Illige2017}. However, these mechanisms could be subject to higher power consumption, oxide breakdown, and/or long switching time determined by the slow migration of oxygen ions \cite{Bi2014, Bauer2014}. Voltage induced transition from ferromagnetic to antiferromagnetic configuration and vice versa has been demonstrated on synthetic antiferromagnetic (SAF) structures using ionic liquid gating \cite{Ming2018}, however, the configuration switches back once the voltage is removed. Similar electric-field induced modulation of the IEC in a SAF layer with an oxide gate has been discussed \cite{JPwangTechcon2018} and combined with the voltage-controlled magnetic anisotropy to demonstrate a bi-directional switching. A different mechanism of manipulating the interlayer exchange coupling has recently been demonstrated \cite{ManginArXiv2019} with an oxide spacer that is capable of exhibiting a metal-insulator transition upon temperature change.

The mechanism proposed in this paper enables bias-dependent IEC via resonant tunneling phenomena, which have the promise to enable lower-voltage controlled operation and faster switching in a non-volatile manner, as compared to the existing mechanisms. Similar resonant tunneling have been demonstrated up to room temperature in a double-barrier magnetic tunnel junction (MTJ) \cite{Liu2013, Tao2015, Suzuki2018}, where the QW forms within a thin magnetic layer and exhibits a bias-dependent oscillation in the magnetoresistance. Here, we argue that the resonant tunneling via a QW formed within a non-magnetic material that exhibits a large IEC and will exploit the property of the material to enable a bias-dependent oscillation in IEC.


\begin{figure}
	\centering
	\includegraphics[width=0.95 \textwidth]{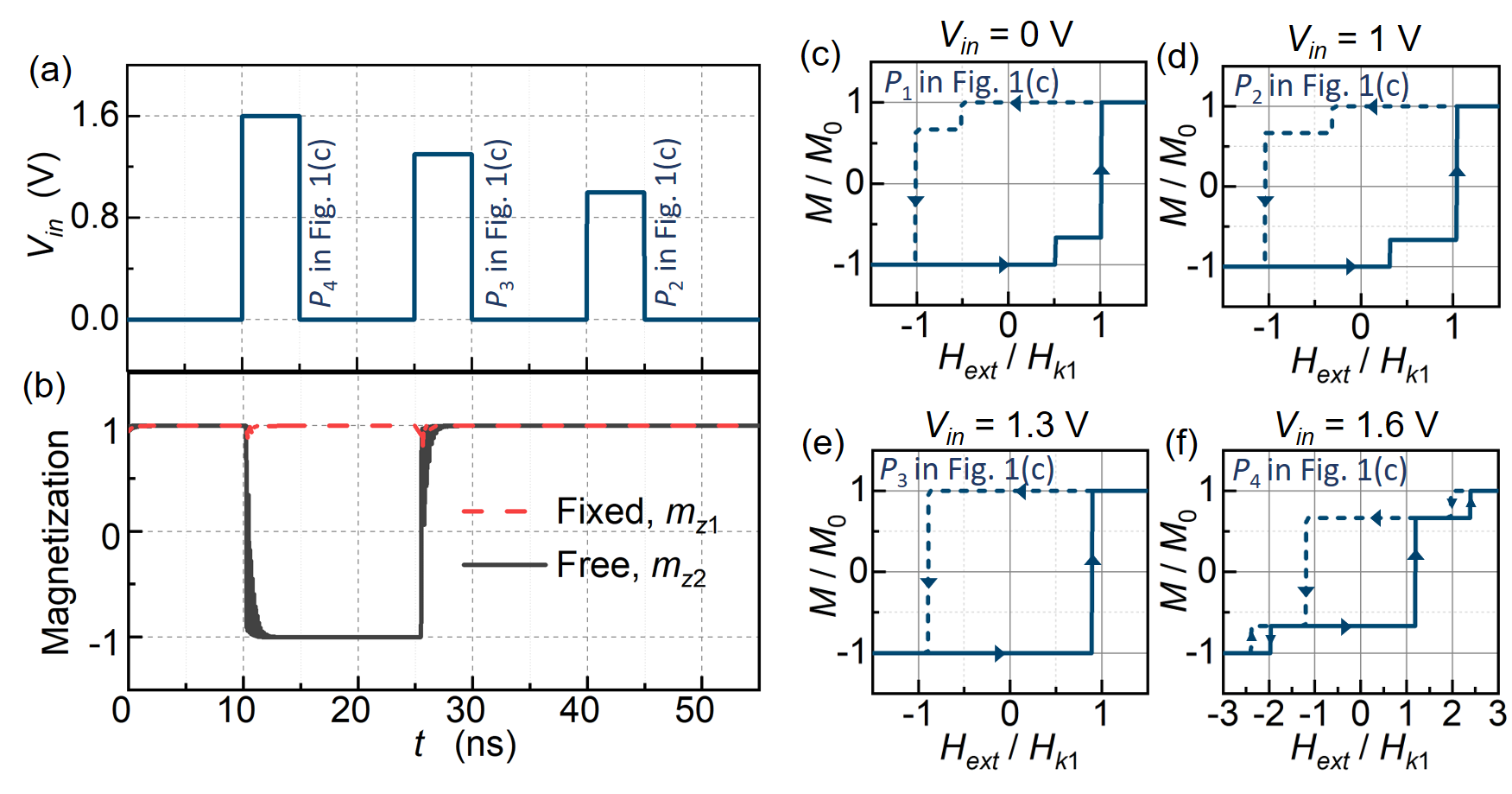}
	\caption{(a) Voltage pulses of same polarity but different magnitudes, applied across the structure in Fig. \ref{1} which induces IEC of different sign and magnitude and switches the magnet if greater than the threshold in Eq. \eqref{switch_th}. (b) $M-H$ loop of the structure in Fig. \ref{1}(c) for different applied voltages.  Fixed magnet: $M_{s1}$ = 1100 emu/cc, $H_{k1}$ = 300 Oe, and $t_{f1}$ = 10 nm. Free magnet: $M_{s2}$ = 1100 emu/cc, $H_{k2}$ = 150 Oe, and $t_{f2}$ = 1.5 nm. Cross-sectional area is $500\times 300$ nm and Gilbert damping is 0.01.}\label{2}
\end{figure}

\section{IEC assisted magnetization dynamics}

\textit{Magnetization switching.} The bias-dependent IEC should switch the free magnet to have either F or AF configuration with respect to the fixed magnet, as dictated by the sign of the IEC. The configuration will be retained once the electric-field is removed because the thick oxide barriers diminish the equilibrium IEC. Such bias-dependent change in IEC sign will enable magnetization switching in both directions for the same voltage polarity (but different magnitudes), which is very different from existing mechanisms for electrical switching. We simulate with three consecutive voltage pulses across the structure, with pulse width of 5 ns each and pulse heights corresponding to P4, P3, and P2 points in Fig. \ref{1}(c), respectively. 

We start with an F configuration as the initial condition and show the $z$ component of the magnetization vectors ($m_{z1}$ for fixed and $m_{z2}$ for free) as a function of time, in Fig. \ref{2}(b). The magnetization dynamics have been simulated with an exchanged coupled LLG model \cite{Victora2005, Camsari2016} assuming single domain magnets under zero external magnetic field and negligible spin current. The first pulse (P4) induces an AF IEC peak and switches the free magnet to make it AF with respect to the fixed magnet. The AF configuration is retained when the pulse is removed. The second pulse (P3) induces a F IEC peak which switches the free magnet back to a F configuration with respect to the fixed magnet. Again, the F configuration is retained when the pulse is removed. The third pulse induces a weak AF peak with strength $|J_{ex}|$ below the switching threshold, hence, the free magnet does not switch.

\textit{Switching threshold.} The threshold of the IEC strength required for magnetization switching ($|J_{ex0}|$) is given by
\begin{equation}
\label{switch_th}
{|J_{ex0}|} \times S  = \dfrac{{2{E_1}{E_2}}}{{{E_1} + {E_2}}},
\end{equation}
Here $S$ is the cross-sectional area of the device, ${E_1}$ and $E_2$ are the thermal energy barriers of the fixed and the free magnets i.e. $E_1 > E_2$. $E_{1,2}$ are determined by $\frac{1}{2}{M_{s}}{H_{k}}{\Omega}$ of the corresponding magnet where $M_s$ is the saturation magnetization, $H_k$ is the anisotropy field, and $\Omega$ is the magnet volume. Eq. \eqref{switch_th} is obtained from the coupled LLG equation in Refs. \cite{Victora2005, Camsari2016} under zero external magnetic field and zero spin current. Note that Eq. \eqref{switch_th} is valid for magnets with both in-plane and perpendicular magnetic anisotropies. An analytical derivation of  Eq. \eqref{switch_th} starting from the LLG model \cite{Victora2005, Camsari2016} and assuming perpendicular anisotropy is provided in the supplementary information. The IEC induced switching mechanism does not depend on the Gilbert damping and the magnetic anisotropy, which is different from the non-equilibrium spin current based switching mechanisms \cite{Sun2000} and similar to an external magnetic field induced switching. 


\textit{Switching time.} The IEC assisted switching time is given by
\begin{equation}
\label{switch_tim}
{t_{sw}} = \dfrac{{2 \pi  }}{{{\alpha _g}\gamma {H_{K2}}}}\dfrac{{|{J_{ex0}}|}}{{|{J_{ex}}|}}\left( {\dfrac{\pi }{2} - {\theta _0}} \right),
\end{equation}
where $\theta _0$ is the initial angle between the fixed and the free magnetizations in the units of radian (rad). $\gamma$ is the gyromagnetic ration in the units of rad s$^{-1}$ T$^{-1}$. Eq. \eqref{switch_tim} is valid only for magnets with perpendicular anisotropies. The constant prefactors are deduced from empirical fitting which can be revisited as the field evolves, however, functional dependence on the parameters are benchmarked with detailed LLG simulations in the supplementary information. The switching time for magnets with in-plane anisotropies are also affected by the demagnetization field and can be analyzed directly using the LLG equation. Note that the switching time is lower for larger IEC energy which is similar to the STT-mechanism where a higher spin current yields a lower switching time \cite{Sun2000}. The pulse rise and fall times in Fig. \ref{2}(a) should be faster than the switching time in Eq. \eqref{switch_tim} in order to avoid any unwanted reverse switching due to slow change in the voltage amplitude.

\begin{figure}
	\centering
	\includegraphics[width=0.9 \textwidth]{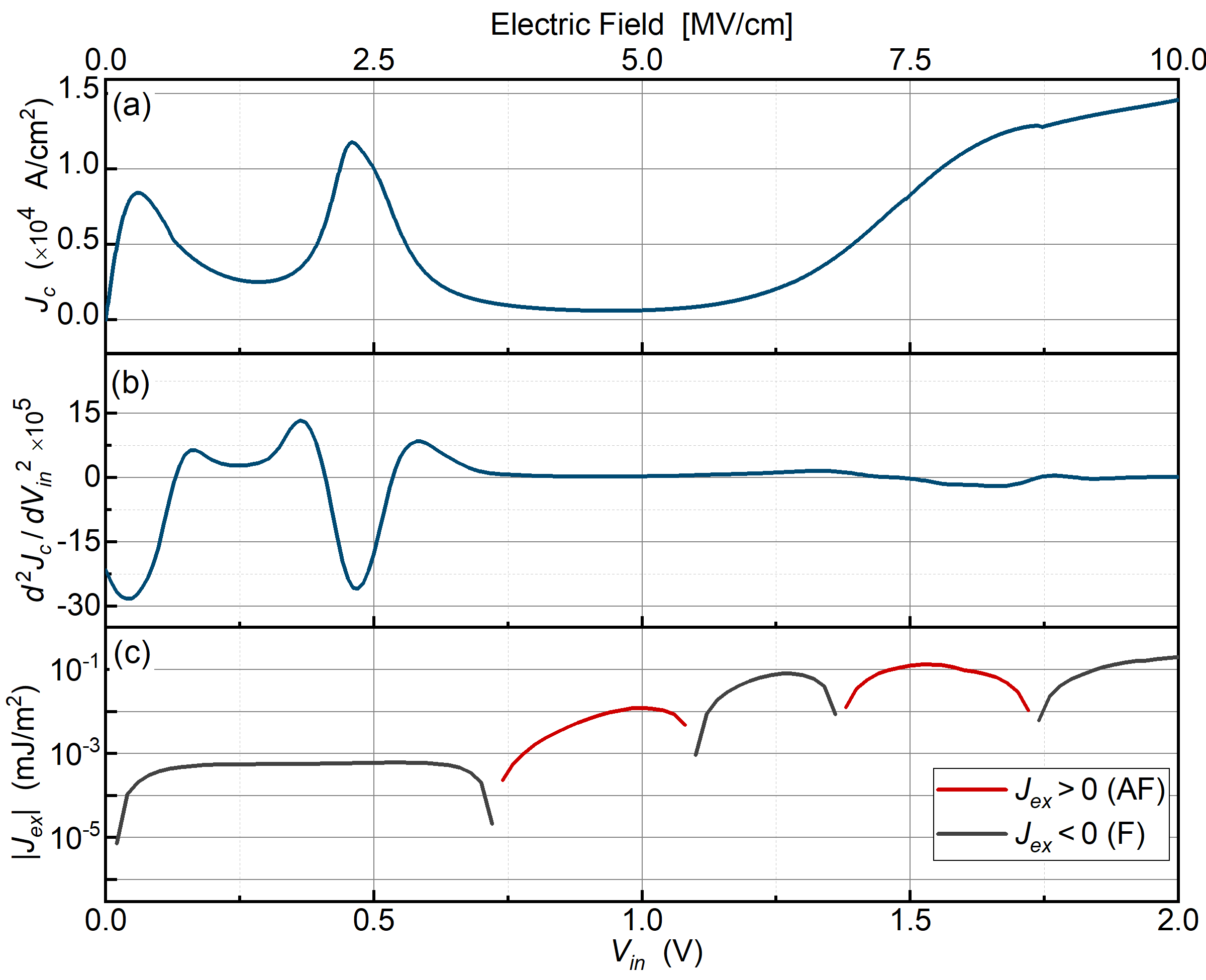}
	\caption{(a) Charge current density $J_c$ of the structure in Fig. \ref{1}(c) as a function of voltage. The current peaks are observed near QW states. (b) $d^2J_c/dV_{in}^2$ as a function of input voltage. (c) The magnitude of the IEC in Fig. \ref{1}(c) in log scale.}\label{3}
\end{figure}

\textit{Signature of bias-dependent IEC.} The oscillation in equilibrium IEC as a function of the distance between the two magnets is observed in spin-valve like structures as a shift in the switching field within the hysteresis loop of the total moment $M$ \cite{Parkin1991,Fullerton1993,Bennett1990}. We argue that similar shift in the switching field should be observed for the proposed structure in Fig. \ref{1}(c) as a function of input voltage, while the distance between the two magnets is kept fixed. We simulate such $M-H$ loops (see Figs. \ref{2}(c)-(f)) using the coupled LLG model \cite{Victora2005,Camsari2016} under an external magnetic field ($H_{ext}$) sweep, assuming negligible spin current, and using the bias-dependent IEC energy $J_{ex}$ calculated from the NEGF. For $V_{in}=0$ V, $J_{ex}$ is negligible (see, P1 in Fig. \ref{1}(c)) and the $M-H$ loop is such that the two magnets are switching at the corresponding coercive fields (here, $H_{k1}=300$ Oe for fixed and $H_{k2}=$ 150 Oe for free in the simulations), as shown in Fig. \ref{2}(c). Note that the $x$-axis of Figs. \ref{2}(c)-(f) are normalized with respect to $H_{k1}$ and $y$-axis are normalized with respect to the total magnetic amoment $M_0=M_1+M_2$ of the structure.

At $V_{in}=1$ V, we observe the first bias-dependent AF peak of $+0.012$ mJ/m$^2$ (see, P2 in Fig. \ref{1}(c)). This exhibits as a sizable lowering of the switching field of the free magnet as shown in Fig. \ref{2}(d). The fixed magnet switching field also changes slightly. Note that such bias-dependent change in the switching field will be an indication of the bias-dependent IEC, even if the IEC strength is below the switching threshold. At $V_{in}=1.3$ V, we observe the subsequent F peak of $-0.08$ mJ/m$^2$ (see, P3 in Fig. \ref{1}(c)) and the $M-H$ loop exhibits a rectangular shape (see Fig. \ref{2}(d)). This suggests that the two magnets are in F configuration and they are switching simultaneously. At $V_{in}=1.6$ V, we observe the second AF peak (see, P4 in Fig. \ref{1}(c)) and the $M-H$ loop exhibits a large split due to a large shift in the free magnet switching field. The middle loop corresponding to the AF configuration with the total moment $M_1-M_2$.

\section{Discussion on efficient design:}

\textit{Delay and energy.} The Gilbert damping and the magnetic anisotropy of the proposed device can be tuned to achieve fast switching operations (see Eq. \eqref{switch_tim}), while the IEC switching threshold $J_{ex0}$ remains unaffected (see Eq. \eqref{switch_th}). This is different from the STT-based devices where there exists a trade-off between the switching threshold and switching time in terms of these parameters \cite{Sun2000}. We argue the proposed structure can achieve a sizable bias-dependent $J_{ex}>J_{ex0}$ while the current density $J_c$ of the structure can be significantly lower. This is because the magnitude of the $J_c$ is limited by the separation between the contact occupation factors $f_1 - f_2$ where $f_{1,2} = 1/\left(1+\exp\left(\left(E-\mu_{1,2}\right)/k_BT\right)\right)$ with $E$ being the energy, $k_B$ being the Boltzmann constant, and $T$ being the temperature. However, with the increased transmission between the two FMs, a sizable IEC is felt by the FMs contributed by all the occupied QW electronic states positioned under the electrochemical potentials. Thus, the proposed device have the potential to achieve significantly low energy-delay product.

We have calculated the current density of the proposed structure using NEGF as $J_c = \int {dE\,{\mathop{\rm Re}\nolimits} \left[ {{\rm{Tr}}\left[ {\frac{i}{\hbar }\left( {H{G^n} - {G^n}H} \right)} \right]} \right]}$, which is shown in Fig. \ref{3}(a). When the contact electrochemical potential is at a QW energy state, the transmission coefficient between the magnets increases. This, in turn, increases the conductance of the structure and the $J_c$ exhibits a peak at the QW state, as shown in Fig. \ref{3}(a). The position of the QW states can also be detected by looking at the sign change in $d^2J_c/dV_{in}^2$, as shown in Fig. \ref{3}(b).

\begin{figure}
	\centering
	\includegraphics[width=1 \textwidth]{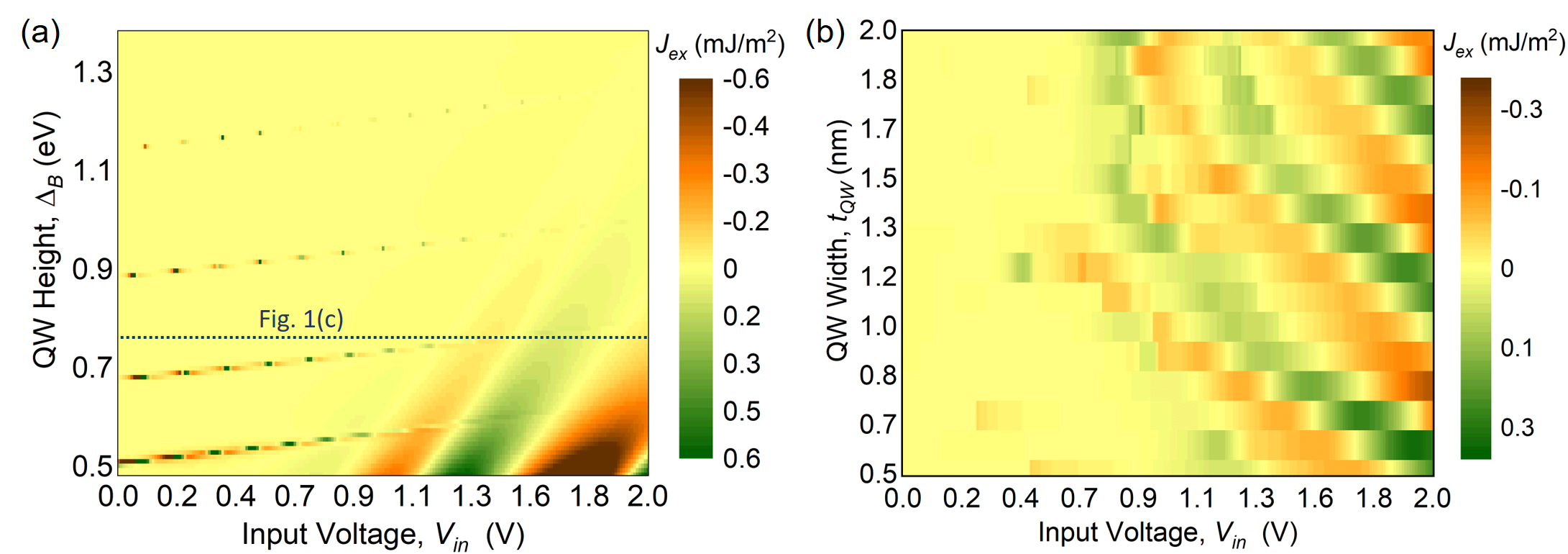}
	\caption{Bias-dependent oscillatory IEC for different (a) barrier heights $\Delta_B$ and (b) QW widths $t_{QW}$. The oscillatory IEC moves toward higher voltage for higher $\Delta_B$ and moves toward lower voltage for wider QWs.}\label{4}
\end{figure}

\textit{QW dimensions.} The charge current in the structure can be further decreased by using a different oxide material with larger barrier height e.g. Al$_2$O$_3$ \cite{Wilt2017}, HfO$_2$ \cite{Shukla_PRB_2017}, SiO$_2$ \cite{Shukla_PRB_2017}, TiO$_2$ \cite{Konenkamp2000}, etc. We have analyzed the effect of various barrier height on the proposed mechanism of inducing a bias-dependent IEC in Fig. \ref{4}(a). A sizable IEC strength persist even for reasonably high barrier heights contributed by the large number of states within the spacer metal, positioned below the Fermi level. However, the bias-dependent oscillatory nature of the IEC shifts toward the higher operating voltage as we increase the barrier height. We also noted from our simulations that for a given barrier height, an additional sharp oscillation peak can be observed at lower operating voltages which occurred due to a small mismatch of transmission coefficients for the parallel and the antiparallel configurations. This sharp oscillation shifts toward higher operating voltages linearly with the increasing barrier height and periodically reoccurs at the lower voltage side. For a given barrier height, the voltage window required to observe this sharp oscillation is very small as compared to the technological interest and further evaluation of this additional oscillation we leave as future work.

We have also analyzed the effect of the spacer metal thickness on the proposed bias-dependent oscillatory IEC mechanism, as shown in Fig. \ref{4}(b). It is interesting to note that the oscillation peaks occur at lower operating voltages for a thicker spacer layer, but the strength of the peaks decreases as well. The former is observed because the spacer layer thickness defines the QW width. An increase in the QW width lowers the discrete energy states as well as the spacing between two consecutive states. Thus the operating voltage to observe an IEC peak lowers with increasing spacer thickness. Similar lowering of QW energy states for wider QW and its consequence on the operating voltage has been discussed for double-barrier MTJ \cite{Tao2015}. The later observation is due to the fact that the distance between two magnets is increasing with increased spacer thickness and the IEC strength weakens. Similar weakening of the IEC strength with increasing distance between the magnets has been discussed for spin-valve like geometries \cite{Parkin1991,Petroff1991,Fullerton1993,Bruno1995,Bruno1995, Haney2009, Slonczewski1989}.


\section{Conclusion:}

In conclusion, we propose a structure that enable  electric-field controlled magnetization switching assisted solely the interlayer exchange coupling (IEC) between the fixed and the free magnets. The two magnets are separated by two oxide barriers sandwiching a metallic spacer that exhibits high IEC strength. The basic idea relies on formation of a quantum-well (QW) within the metallic spacer, which contains discrete energy states above the Fermi level exhibiting high transmission coefficients due to the resonant tunneling phenomena. When the contact electrochemical potential is at one of these discrete states, the two magnets feel a sizable IEC contributed by all the filled QW states under the electrochemical potential. We predict a bias-dependent oscillation in IEC that grows by magnitude with voltage. Such oscillatory IEC can enable bi-directional switching for the same voltage polarity but different magnitudes above the switching threshold value. We show that the switching threshold of this new mechanism is independent of the Gilbert damping and same for magnets with in-plane and perpendicular anisotropies. However, the switching time is inversely proportional to the Gilbert damping and different for magnets with in-plane and perpendicular anisotropies. This decoupling of the dynamic parameter may lead to significant lowering of the energy-delay product as compared to the state-of-the-art mechanisms.

\begin{addendum}
 \item This work was in part supported by ASCENT, one of six centers in JUMP, a SRC program sponsored by DARPA and in part by the Center for Energy Efficient Electronics Science (E3S), NSF Award 0939514.
\end{addendum}

\appendix

\section{NEGF Model}
In this section, we discuss the Non-Equilibrium Green's Function (NEGF) method \cite{Datta1995,Datta2012} used for quantum-transport simulations on the structure in Fig. 1 in the main manuscript.

\subsection{Hamiltonian}

We write the tight-binding Hamiltonian of the structure as the following
\begin{equation}
\label{schrod}
{\left[ \beta  \right]^\dag }\Psi (n - 1) + \left[ \alpha  \right]\Psi (n) + \left[ \beta  \right]\Psi (n + 1) = E\Psi (n),
\end{equation}
where $\Psi$ is the wave function of the $n^\text{th}$ lattice point along $y$-direction. We work with a single band effective mass Hamiltonian,  described by (1) equilibrium electrochemical potential $\mu_{eq}$, (2) exchange splitting $\Delta_{ex}$, (3) barrier heights, $\Delta_B$, (4) well-depth, $\Delta_{B0}$, (5) effective mass for ferromagnet $m^*_{f\uparrow}=m^*_{f\downarrow}=m^*_{f}$, (6) effective mass for oxide $m^*_{ox}$, (7) effective mass for metallic spacer $m^*_{n}$, and (8) contact electrochemical potentials $\mu_{1,2}$. Note that we view these parameters to take into account wide varieties of factors like imperfection at the ferromagnet-non-magnet interfaces by assuming effective values. Below, we present Hamiltonian for each transverse mode with wave vector $k_{\parallel}$ in the device.

\subsubsection{Ferromagnetic Layers}

\begin{figure}
	\centering
	\includegraphics[width=0.8 \textwidth]{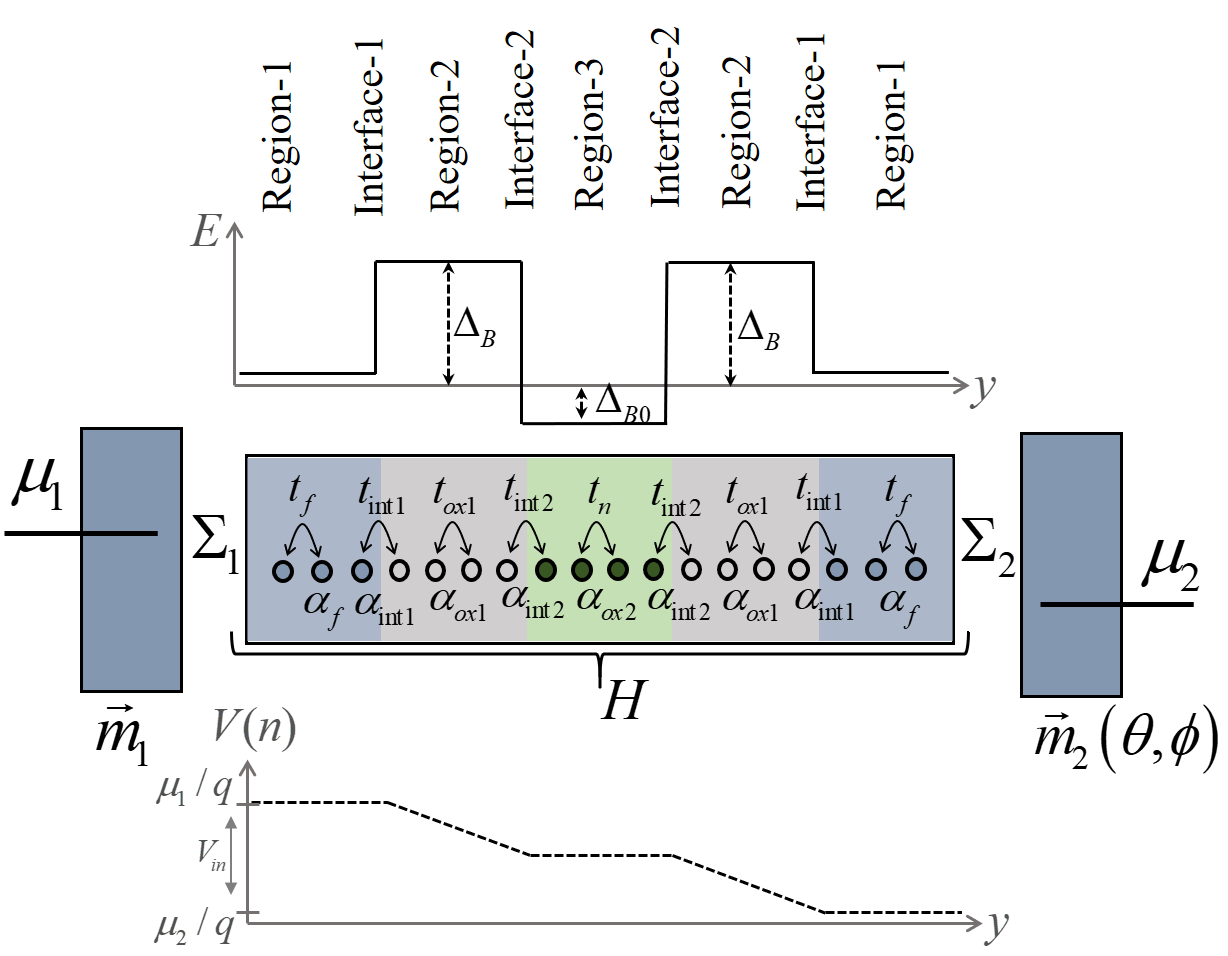}
	\caption{NEGF setup for the structure in Fig. 1 of the main manuscript.}\label{S1}
\end{figure}

For the lattice points within the ferromagnetic layers (indicated by region-1 in Fig. \ref{S1}), we have
\begin{subequations}
	\begin{equation}
	\label{alph_FM}
\left[ \alpha  \right] = \left( {2{t_f} + \frac{{{\hbar ^2}k_\parallel ^2}}{{2m_f^*}} + qV\left( n \right)} \right){I_{2 \times 2}} + \frac{{{\Delta _{ex}}}}{2}\left( {I - \vec \sigma  \cdot \vec m} \right),
	\end{equation}
	\begin{equation}
\left[ \beta  \right] = -2{t_f}{I_{2 \times 2}},
	\end{equation}
\end{subequations}
with $t_f=\dfrac{\hbar^2}{2m_f^*a^2}$, $q$ is the electron charge, $\hbar$ is the reduced Planck's constant, $m_f^*$ is the effective mass in the ferromagnet, $k_\parallel$ is the transverse wavevector, $\Delta _{ex}$ is the exchange-splitting energy in the ferromagnet, $I_{2 \times 2}$ is a $2 \times 2$ identity matix, $ \vec \sigma $ is the Pauli spin matrices, and $a$ is the lattice distance. Here, $qV(n)$ is the potential on the $n^\text{th}$ lattice point of the structure. The potential is varied by the applied voltage $V_{in}$ across the structure which drops across each layer according to the resistance of each layer, as shown in Fig. \ref{S1}. $\vec{m}_1$ and $\vec{m}_2$ are the magnetization vector of the fixed and the free magnet, respectively. Note that in Eq. \eqref{alph_FM}, $\vec{m}=\vec{m}_1$ for fixed magnet region and $\vec{m}=\vec{m}_2$ for free magnet region. Here, $\vec{m} \cdot \vec{M} = \cos\theta$.

\subsubsection{Oxide Layers}

For the lattice points within the oxide layers (indicated by regions 2 in Fig. \ref{S1}), we have
\begin{subequations}
	\begin{equation}
	\left[ \alpha  \right] = \left( {2{t_{ox}} + \frac{{{\hbar ^2}k_\parallel ^2}}{{2m_{ox}^*}}+\Delta_{B} + qV\left( n \right)} \right){I_{2 \times 2}},
	\end{equation}
	\begin{equation}
	\left[ \beta  \right] = -2{t_{ox}}{I_{2 \times 2}},
	\end{equation}
\end{subequations}
with $t_{ox}=\dfrac{\hbar^2}{2m_{ox}^*a^2}$, $m_{ox}^*$ is the effective mass, and $\Delta_{B}$ is the barrier height.

\subsubsection{Spacer Layer}

For the lattice points within the metallic spacer layer (indicated by region 3 in Fig. \ref{S1}), we have
\begin{subequations}
	\begin{equation}
	\left[ \alpha  \right] = \left( {2{t_{n}} + \frac{{{\hbar ^2}k_\parallel ^2}}{{2m_{n}^*}}-\Delta_{B0} + qV\left( n \right)} \right){I_{2 \times 2}},
	\end{equation}
	\begin{equation}
	\left[ \beta  \right] = -2{t_{n}}{I_{2 \times 2}},
	\end{equation}
\end{subequations}
with $t_{n}=\dfrac{\hbar^2}{2m_{n}^*a^2}$, $m_{n}^*$ is the effective mass in the spacer, and $\Delta_{B0}$ is the depth of the spin-dependent QW like potential below the equilibrium electrochemical potential.

\subsubsection{Ferromagnet $|$ Oxide Interfaces}
For the lattice points for the interface of the ferromagnet and oxide-1 layers (indicated by interface-1 in Fig. \ref{S1}), we have
\begin{subequations}
	\begin{equation}
	\begin{array}{l}
\left[ \alpha  \right] = \left( {{t_f}  + {t_{ox}} + \dfrac{{{\hbar ^2}k_\parallel ^2}}{{4m_f^*}}+ \dfrac{{{\hbar ^2}k_\parallel ^2}}{{4m_{ox}^*}} +\dfrac{\Delta_{B}}{2} + {qV\left( n \right)}} \right){I_{2 \times 2}} \\\;\;\;\;\;\;\;\;\;\;\;\;\;\;\;\;\;\;\;\;\;\;\;\;\;+ \dfrac{{{\Delta _{ex}}}}{4}\left( {I - \vec \sigma  \cdot \vec m} \right),
	\end{array}	
	\end{equation}
	\begin{equation}
	\left[ \beta  \right] = 0,
	\end{equation}
\end{subequations}
where $\vec m \equiv \vec m_1$ is for interface 1 and $\vec m \equiv \vec m_2$ is for interface 2.

\subsubsection{Oxide $|$ Spacer Interfaces}
For the lattice points for the interface of the oxide and spacer layers (indicated by interface-2 in Fig. \ref{S1}), we have
	\begin{equation}
	\begin{array}{l}
	\left[ \alpha  \right] = \left( {{t_{ox1}} + {t_{n}} + \dfrac{{{\hbar ^2}k_\parallel ^2}}{{4m_{ox1}^*}} + \dfrac{{{\hbar ^2}k_\parallel ^2}}{{4m_{n}^*}} } \right){I_{2 \times 2}}\\\;\;\;\;\;\;\;\;\;\;+ \left( { \dfrac{{{\hbar ^2}k_\parallel ^2}}{{4m_{n}^*}} +\dfrac{\Delta_{B}+\Delta_{B0}}{2} + {qV\left( n \right)}} \right){I_{2 \times 2}}.
	\end{array}
	\end{equation}

\subsection{Self-Energy of Contacts}
We will present self-energy matrices for each transverse mode with wave vector $k_{\parallel}$ in the device. The self-energy matrices are given by
\begin{equation}
\label{self_energ}
{\Sigma _{1,2}} =  \left[ {\begin{array}{*{20}{c}}
	- {t_f}{{e^{ik_{1,2}^ \uparrow a}}}&0\\
	0&- {t_f}{{e^{ik_{1,2}^ \downarrow a}}}
	\end{array}} \right],
\end{equation}
where $k_{1,2}^ \uparrow$ and $k_{1,2}^ \downarrow$ are longitudinal wavevectors of up and down spins respectively, which are estimated from
\begin{subequations}
	\begin{equation}
	E = E_{C,1}^{ \uparrow , \downarrow } + \frac{{q{V_{in}}}}{2} + \frac{{{\hbar ^2}k_\parallel ^2}}{{2m_f^*}} + 2{t_f}\left[ {1 - \cos \left( {k_1^{ \uparrow , \downarrow }} \right)} \right],
	\end{equation}
	\begin{equation}
	E = E_{C,2}^{ \uparrow , \downarrow } - \frac{{q{V_{in}}}}{2} + \frac{{{\hbar ^2}k_\parallel ^2}}{{2m_f^*}} + 2{t_f}\left[ {1 - \cos \left( {k_2^{ \uparrow , \downarrow }} \right)} \right],
	\end{equation}
\end{subequations}
with $E_{C,1}$ and $E_{C,2}$ being the bottom of the conduction bands for left and right magnetic contacts.


Note that in the present discussion, the magnetization $\vec{m}_2$ of the free magnet can lie in an arbitrary direction, hence, the effective $\Sigma _{2}$ is given by
\begin{equation}
\label{self_energ2}
{\Sigma _{2}} =\Re\left(\theta,\phi \right) \left[ {\begin{array}{*{20}{c}}
	 - {t_f}{{e^{ik_{2}^ \uparrow a}}}&0\\
	0& - {t_f}{{e^{ik_{2}^ \downarrow a}}}
	\end{array}} \right] \Re^\dagger \left(\theta,\phi \right),
\end{equation}
with $\Re\left(\theta,\phi \right)$ being a rotational matrix given by
\begin{equation}
\Re\left(\theta,\phi \right)=\left[ {\begin{array}{*{20}{c}}
	{\cos \dfrac{\theta }{2}{e^{ - i\dfrac{\phi }{2}}}}&{ - \sin \dfrac{\theta }{2}{e^{ - i\dfrac{\phi }{2}}}}\\
	{\sin \dfrac{\theta }{2}{e^{ + i\dfrac{\phi }{2}}}}&{\cos \dfrac{\theta }{2}{e^{ + i\dfrac{\phi }{2}}}}
	\end{array}} \right].
\end{equation}
Note that in the analysis presented here, we assume that the magnetization $\vec{m}_2$ lies in the $z$-$x$ plane creating an angle $\theta$ with the $\vec{m}_1$ with $\phi=0$.

\subsection{NEGF Quantities}
We calculate the following quantities:

\begin{itemize}
	\item Green's function:
	\begin{equation}
	G^R=\left[EI-H-\Sigma\right]^{-1},
	\end{equation}
	with $\Sigma=\Sigma_1+\Sigma_2$ (see Eqs. \eqref{self_energ} and \eqref{self_energ2}). $H$ is the Hamiltonian of the system as discussed earlier.
	
	\item Spectral function:
	\begin{equation}
	\label{spec_func}
	A=G^R\,\Gamma\, \left(G^R\right)^\dagger,
	\end{equation}
	with $\Gamma=\Gamma_1+\Gamma_2$ and $\Gamma_{1,2}$ are broadening functions which represent the anti-Hermitian part of $\Sigma_{1,2}$ i.e. $\Gamma_{1,2}=i\left(\Sigma_{1,2}-\Sigma_{1,2}^\dagger\right)$. Note that $A/2\pi$ provides the density of states of the system.
	
	\item Correlation function:
	\begin{equation}
	\label{e_func}
	G^n=G^R \Sigma^{in} \left(G^R\right)^\dagger,
	\end{equation}
	with $\Sigma^{in}=\Sigma^{in}_1+\Sigma^{in}_2$ being the in-scattering function. Note that $G^n/2\pi$ provides the electron density of the system.
	
	\item In-scattering function:
	\begin{equation}
	\Sigma^{in}_{1,2}=\Gamma_{1,2}f_{1,2}.
	\end{equation}
	with $f_{1,2}$ being the Fermi occupation factors of contacts 1 and 2, given by
	\begin{equation}
	\label{f_func}
	f_{1,2}=\dfrac{1}{1+\exp\left({\dfrac{E-\mu_{1,2}}{k_BT}}\right)}.
	\end{equation}
	Here, $\mu_{1,2}$ are electrochemical potentials of contacts 1 and 2, $k_B$ is the Boltzmann constant, and $T$ is the temperature. Note that in the present discussion: $qV_{in}=\mu_1 - \mu_2$.
	
	\item Current operator:
	\begin{equation}
	I^{op}=\dfrac{i}{\hbar}\left(H{G^n} - {G^n}H\right).
	\end{equation}
	Current operator between two adjacent lattice points $j$ and $j+1$ is given by
	\begin{equation}
	I^{op}_{j,j+1}=\dfrac{i}{\hbar}\left(H_{j,j+1}{G^n_{j,j+1}} - {G^n_{j,j+1}}H_{j,j+1}\right).
	\end{equation}
	The charge current density is given by
	\begin{equation}
    J_c=\int dE\,\text{Re}\left(\text{Tr}\left(I^{op}_{j,j+1}\right)\right).
	\end{equation}
	The spin current density is given by
	\begin{equation}
	\label{spin_cur}
	\vec{J}_s=\int dE\,\text{Re}\left(\text{Tr}\left(\vec{\sigma} I^{op}_{j,j+1}\right)\right),
	\end{equation}
	where $\vec{\sigma}$ is the Pauli spin matrices.
\end{itemize}

\begin{figure*}
	\centering
	\includegraphics[width=0.9 \textwidth]{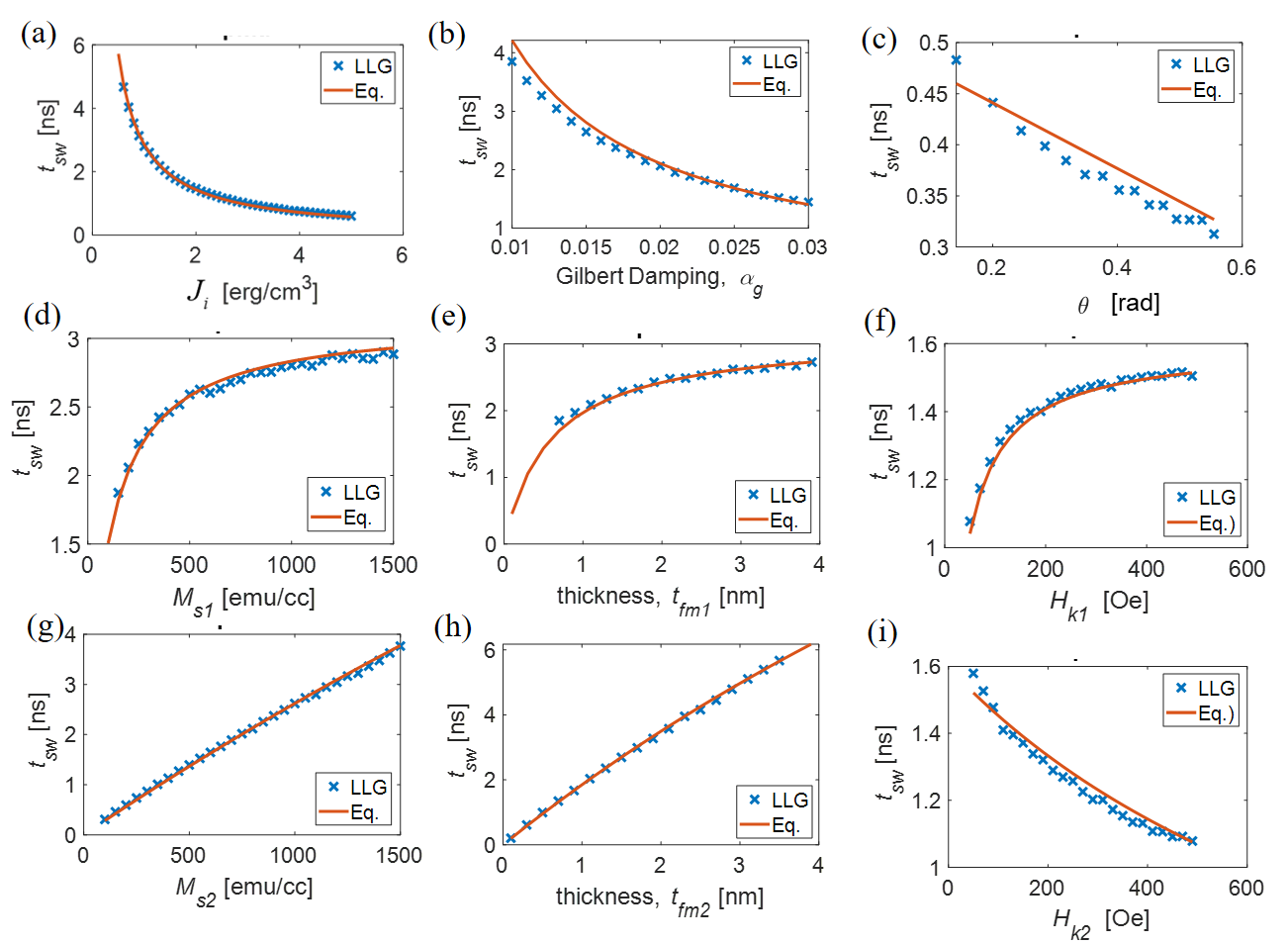}
	\caption{Comparison of the simple expression of switching time ($t_{sw}$) in Eq. \eqref{switch_tim} with detailed numerical simulation using coupled-LLG in Eqs. \eqref{LLG1}-\eqref{LLG2}. (a) $t_{sw}$ vs. applied IEC $J_{ex}$. (b) $t_{sw}$ vs. Gilbert damping $\alpha_g$. (c) $t_{sw}$ vs. initial angle of the free magnet $\theta_0$. (d) $t_{sw}$ vs. saturation magnetization of the fixed magnet $M_{s1}$. (e) $t_{sw}$ vs. thickness of the fixed magnet $t_{fm1}$. (f) $t_{sw}$ vs. anisotropy field of the fixed magnet $H_{K1}$. (g) $t_{sw}$ vs. saturation magnetization of the free magnet $M_{s2}$. (h) $t_{sw}$ vs. thickness of the free magnet $t_{fm2}$. (i) $t_{sw}$ vs. anisotropy field of the free magnet $H_{K2}$.}\label{S2}
\end{figure*}

\subsection{IEC Calculation}

We start by calculating the $z$ spin density at an interfaces from NEGF parameters calculated based on the 1D tight-binding Hamiltonian as
\begin{equation}
    n_{z,j}^{1D}=\text{Re}\left(\text{Tr}\left({\sigma_z} G^n_{j,j+1}\right)\right),
\end{equation}
where $\sigma_z$ is the $z$-Pauli matrix, and $G^n$ is the correlation function. We assume that the structure under consideration is spatially uniform along the transverse directions and transverse modes are nearly decoupled so that transport can be analyzed with 1D Hamiltonian for every mode. Under such assumption, we do a mode summation on the transverse plane ($y$-$z$ plane), to estimate the 2D $z$ spin density at $j^{\text{th}}$ lattice point, as given by
\begin{equation}
\label{IEC}
n_s\left(x_j\right) = \frac{1}{2\pi}\int D_0\; n_{z,j}^{1D} \;d\epsilon_T,
\end{equation}
where $\epsilon_T=\dfrac{\hbar^2 \left(k_y^2+k_z^2\right)}{2m^*}$ is the energy along the transverse plane $y$-$z$, and $D_0$ is the 2D density of states on the transverse plane. 

We approximate the $D_0$ with the 2D density of states of the bulk given by \cite{Datta2012}
\begin{equation}
 D_0 = \dfrac{m^*}{\pi\hbar^2},   
\end{equation}
where $m^*$ is the effective mass and $\hbar=h/2\pi$. The change in $z$ spin density across the quantum-well is given by
\begin{equation}
    \Delta n_s = n_s\left(x_j = 0\right)-n_s\left(x_j = t_{QW}\right).
\end{equation}

The total change in energy of the occupied states across the quantum-well is calculated as
\begin{equation}
\label{change_energy}
    \Delta E=\displaystyle\int_{-\infty}^{+\infty} E \; \Delta n_s\left(E\right)\;dE.
\end{equation}
which is in the units of J-m$^{-2}$.

In a magnetic structure under consideration, the interlayer exchange coupling energy density is calculated as the change in Eq. \eqref{change_energy} from ferromagnetic (F) to antiferromagnetic (AF) configurations \cite{Bruno1995, Haney2009, Slonczewski1989}
\begin{equation}
\label{IEC}
{J_{ex}} = \Delta E_F - \Delta E_{AF}.
\end{equation}
We have calibrated the spacer parameters such that the first AF IEC strength is around $\sim4$ mJ/m$^2$, as shown in Fig. \ref{1}(a). For simplicity, we assume two oxide barriers are same with width of 1 nm each. The oxide barrier heights were set around $0.7\sim0.8$ eV, as typically observed for MgO \cite{Schleicher2014, Yuasa2004}. We assume that the voltage applied across the two-terminal structure mostly drops across the two oxide barriers. The NEGF setup has been discussed in detail in the supplementary information.

\section{Switching Threshold and Time}

In this section, we will derive the IEC assisted switching threshold in Eq. (2) of the main manuscript. We will also discuss the switching time in Eq. (3) in terms of detailed numerical simulation using a coupled LLG equation.

\subsection{LLG Equation}

We assume mono-domain magnets and analyze the magnetization dynamics using a coupled-Landau-Lifshitz-Gilbert (LLG) equation \cite{Victora2005, Camsari2016}, given by
\begin{subequations}
\begin{equation}
\label{LLG1}
\begin{array}{l}
\left( {1 + \alpha _g^2} \right)\dfrac{{d\hat m_1}}{{dt}} =  - \gamma \,\hat m_1 \times {{\vec H}_{eff}} - {\alpha _g}\gamma \,\hat m_1 \times \hat m_1 \times {{\vec H}_{eff}}\\
\;\;\;\;\;+ \hat m_1 \times \dfrac{{{{\vec I}_{S}}}}{{q{N_{S,1}}}} \times \hat m_1 + {\alpha _g}\hat m_1 \times \dfrac{{{{\vec I}_{S}}}}{{q{N_{S,1}}}}\\
\, - \,\hat m_1 \times \dfrac{{2{J_{ex}}{S_{tot}}}}{{\hbar {N_{S,1}}}}\hat m_2 - {\alpha _g}\,\hat m_1 \times \hat m_1 \times \dfrac{{2{J_{ex}}{S_{tot}}}}{{\hbar {N_{S,1}}}}\hat m_2,
\end{array}
\end{equation}
\begin{equation}
\label{LLG2}
\begin{array}{l}
\left( {1 + \alpha _g^2} \right)\dfrac{{d\hat m_2}}{{dt}} =  - \gamma \,\hat m_2 \times {{\vec H}_{eff}} - {\alpha _g}\gamma \,\hat m_2 \times \hat m_2 \times {{\vec H}_{eff}}\\
\;\;\;\;\;+ \hat m_2 \times \dfrac{{{{\vec I}_{S}}}}{{q{N_{S,2}}}} \times \hat m_2 + {\alpha _g}\hat m_2 \times \dfrac{{{{\vec I}_{S}}}}{{q{N_{S,2}}}}\\
\, - \,\hat m_2 \times \dfrac{{2{J_{ex}}{S_{tot}}}}{{\hbar {N_{S,2}}}}\hat m_1 - {\alpha _g}\,\hat m_2 \times \hat m_2 \times \dfrac{{2{J_{ex}}{S_{tot}}}}{{\hbar {N_{S,2}}}}\hat m_1,
\end{array}
\end{equation}
\end{subequations}
where $\vec m_1$ and $\vec m_2$ are the magnetization vectors of FM-1 and FM-2 respectively, $\hbar$ is the reduced Planck's constant, $S_{tot}$ is the total cross-sectional area of the magnets, $\gamma$ is the gyromagnetic ratio, $\alpha_g$ is the Gilbert damping constant,  $\vec H_{eff}$ indicates the effective magnetic field which includes the demagnetizing field and the anisotropy field of the corresponding magnet, $J_{ex}$ is given by Eq. (1) in the main manuscript, and $\vec{I}_S$ is given by Eq. \eqref{spin_cur}. $N_{S,1}$ and $N_{S,2}$ are the number of spins in FM-1 and FM-2 respectively, where $N_{S,1/2}=M_{s,1/2}\Omega_{1/2}/\mu_B$, $M_{s1}$ and $M_{s2}$ are saturation magnetizations of FM-1 an FM-2, $\Omega_1$ and $\Omega_2$ are the volumes of FM-1 and FM-2, and $\mu_B$ is the Bohr magneton.

\subsection{Simplification}

We assume that the spin current in the structure is negligible i.e. $\vec{I}_S = 0$. There is no external magnetic field i.e. $\vec{H}_{ext}=0$ and we assume perpendicular magnetic anisotropy so that the effective magnetic field in the structure is given by ${\vec H_{eff1,2}} = {H_{K1,2}}\hat z$. Thus, Eqs. \eqref{LLG1}-\eqref{LLG2} reduces to
\begin{subequations}
	\begin{equation}
	\label{rLLG1}
	\begin{array}{l}
	\left( {1 + \alpha _g^2} \right)\dfrac{{d{m_{z1}}}}{{dt}} = {\alpha _g}{\mkern 1mu} \gamma {H_{K1}}\left( {{\mkern 1mu} {m_{z1}} + \dfrac{{{J_{ex}}S}}{{{E _1}}}\;{m_{z2}}} \right)\left( {1 - m_{z1}^2} \right),
	\end{array}
	\end{equation}
	\begin{equation}
	\label{rLLG2}
	\left( {1 + \alpha _g^2} \right)\dfrac{{d{m_{z2}}}}{{dt}} = {\alpha _g}\gamma {H_{K2}}\left( {{\mkern 1mu} {m_{z2}} + \dfrac{{{J_{ex}}S}}{{{E _2}}}\;{m_{z1}}} \right)\left( {1 - m_{z2}^2} \right).
	\end{equation}
\end{subequations}

\subsection{Derivation of Switching Threshold}

For parallel conditions: $m_{z1}=1-\delta_1$ and $m_{z2}=1-\delta_2$ or $m_{z1}=-1+\delta_1$ and $m_{z2}=-1+\delta_2$, we have from Eqs. \eqref{rLLG1}-\eqref{rLLG2}
\[ - \left( {1 + \alpha _g^2} \right)\frac{{d{\delta _1}}}{{dt}} \approx {\alpha _g}{\mkern 1mu} \gamma {H_{K1}}\left( {{\mkern 1mu} 1 + \dfrac{{{J_{ex}}S}}{{{E _1}}}} \right) {\delta _1}\]
\[ - \left( {1 + \alpha _g^2} \right)\frac{{d{\delta _2}}}{{dt}} \approx{\alpha _g}\gamma {H_{K2}}\left( {{\mkern 1mu} 1 + \dfrac{{{J_{ex}}S}}{{{E _2}}}} \right){\delta _2}\]
which are stable only if $\left(-J_{ex}\times S \right)<E_1$ and $\left(-J_{ex}\times S \right)<E_2$. Note that $\delta_{1,2}\rightarrow0$.

Similarly, for anti-parallel conditions: $m_{z1}=1-\delta_1$ and $m_{z2}=-1+\delta_2$ or $m_{z1}=-1+\delta_1$ and $m_{z2}=1-\delta_2$, we have from Eqs. \eqref{rLLG1}-\eqref{rLLG2}
\[ - \left( {1 + \alpha _g^2} \right)\frac{{d{\delta _1}}}{{dt}} \approx {\alpha _g}{\mkern 1mu} \gamma {H_{K1}}\left( {{\mkern 1mu} 1 - \dfrac{{{J_{ex}}S}}{{{E _1}}}} \right) {\delta _1}\]
\[  \left( {1 + \alpha _g^2} \right)\frac{{d{\delta _2}}}{{dt}} \approx{\alpha _g}\gamma {H_{K2}}\left( {{\mkern 1mu} -1 + \dfrac{{{J_{ex}}S}}{{{E _2}}}} \right){\delta _2}\]
which are stable only if $\left(J_{ex}\times S \right)<E_1$ and $\left(J_{ex}\times S \right)<E_2$.

Thus, the conditions for stability in either parallel or anti-parallel configurations are given by
\begin{subequations}
	\begin{equation}
	\dfrac{{\left| {{J_{ex}}S} \right|}}{{{E _1}}} < 1,
	\end{equation}
	\begin{equation}
	\dfrac{{\left| {{J_{ex}}S} \right|}}{{{E _2}}} < 1,
	\end{equation}
\end{subequations}
which yields the condition for stability as
\begin{equation}
\dfrac{{\left| {{J_{ex}}S} \right|}}{{{E _1}}}+\dfrac{{\left| {{J_{ex}}S} \right|}}{{{E _2}}} < 2.
\end{equation}

Thus, the condition required to make any stable configuration unstable is given by
\begin{equation}
\dfrac{{\left| {{J_{ex}}S} \right|}}{{{E _1}}}+\dfrac{{\left| {{J_{ex}}S} \right|}}{{{E _2}}} \geq 2,
\end{equation}
which in turn, gives the switching threshold expression in Eq. (2) of the main manuscript. Note that for simplicity, the expression for switching threshold is derived assuming perpendicular magnetic anisotropy. However, the expression is valid for in-plane magnetic anisotropy as well.

\subsection{Switching Time}

We provide an expression for the switching time given by Eq. (3) in the main manuscript
\begin{equation*}
\label{switch_tim}
{t_{sw}} = \dfrac{2\pi}{{{\alpha _g}\gamma {H_{K2}}}}\dfrac{{{|J_{ex0}|}}}{{{|J_{ex}|}}}\left( {\dfrac{\pi }{2} - {\theta _0}} \right).
\end{equation*}
We have compared this simple expression with the detailed LLG simulation results from Eqs. \eqref{LLG1}-\eqref{LLG2} as shown in Fig. \ref{S2} for different parameters. In most of our simulations, we have used $M_{s1}=M_{s2}=1100$ emu/cc, $H_{K1}=260$ Oe, $H_{K2}=100$ Oe, $S=$ 500 nm $\times$ 200 nm, $t_{fm1}=10$ nm, $t_{fm2}=1.6$ nm, and $\alpha_g = 0.008$.

\bibliographystyle{naturemag}
\bibliography{Ref}

\end{document}